\documentstyle[12pt]{article}

\begin{document}

\author{L. Herrera\thanks{%
Also at Departamento de F\'\i sica, Facultad de Ciencias,
Universidad
Central de Venezuela, Caracas, Venezuela; email:lherrera@gugu.usal.es}
, A. Di Prisco$^*$ \\
\'Area de F\'\i sica Te\'orica\\
Facultad de Ciencias\\
Universidad de Salamanca\\
37008, Salamanca, Espa\~na.\\
and \and J. Mart\'\i nez \\
Grupo de F\'\i sica Estad\'\i stica\\
Departamento de F\'\i sica\\
Universidad Aut\'onoma de Barcelona\\
08193 Bellaterra, Barcelona, Espa\~na.}
\title{Non spherical sources of strong gravitational fields out of
hydrostatic
equilibrium}
\maketitle

\begin{abstract}
We describe the departure from equilibrium of matter distributions
representing sources for a class of Weyl metric. It is shown that, for
extremely high gravitational fields, slight deviations from spherical
symmetry may enhance the stability of the system weakening thereby its
tendency
to a catastrophic collapse. For critical values of surface gravitational
potential, in contrast with the exactly spherically symmetric case,
the speed of entering the collapse regime decreases substantially
, at least for specific cases.
\end{abstract}

\section{Introduction}

\begin{quote}
``{\sc L'interviewer.} - mais croyez-vous que vous devez \^{e}tre d'accord
avec ce que la plupart de gens autour de vous pensent?''
\\
``{\sc Ruth.} - eh bien, c'est-\`{a}-dire que, quand je ne le suis pas, je
me retrouve toujours \`{a} l'h\^{o}pital...''\\
{\sl R. D. Laing, A. Esterson,} {\it L'\'{e}quilibre mental, la folie et la
famille,} {\sl Ed. Maspero (Paris, 1971).\\}
\end{quote}

As it is well known, since the early seminal work of Israel \cite{1},
the
only static and asymptotically-flat vacuum space-time possessing a
regular
horizon is the Schwarzschild solution. For all the others Weyl exterior
solutions \cite{2}, the physical components of the Riemann tensor
exhibit
singularities at $r=2m$.

Since these components represent the true observables of the theory as
it
appears from the definition of tidal forces \cite{3}, it is intuitively
clear that for high gravitational fields, the evolution of sources of
Weyl
space-time should drastically differ from the evolution of spherical
sources \cite{Bel}. It is important to keep in mind that the sharp difference in
the
behaviour of both types of sources (for very high gravitational fields)
will
exist independently on the magnitude of multipole moments (higher than
monopole)
of the Weyl source. This is so because, as the source approaches the
horizon, any finite perturbation of the Schwarzschild space-time becomes
fundamentally different from any Weyl solution, even when the latter is
characterized by parameters whose values are arbitrarily close to those
corresponding to the spherical symmetry. This point has been stressed
long
time ago \cite{4}, but usually it has been overlooked.

It is the purpose of this work to study the evolution of axisymmetric
sources for very high gravitational fields. This will allow us to put
in
evidence the role played by the non sphericity (however small) of the source,
on the outcome of evolution. However, instead of following the
evolution
of the system long time after its departure from equilibrium, what would
require the use of numerical procedures, we shall evaluate the source
immediately after such departure. Here ``immediately'' means on a time
scale
smaller than hydrostatic time scale -see section 4 for more details. In
doing so we shall avoid the introduction of numerical procedures.
On the other hand, however, we
shall
obtain only indications about the tendency of the object and not a
complete
description of its evolution. In spite of this limitation, this approach
has
proven to be useful in the study of spherically symmetric case (see
\cite{5}
and \cite{6} and references therein).

As initial configurations, we shall consider two interior metrics. These
ones
were found some years ago by Stewart {\it et al.} \cite{7}, following a
prescription given by Hern\'{a}ndez \cite{8} allowing to
obtain interior solutions of Weyl space time, from known spherically
symmetric interior solutions.

The configurations to be considered are sources of the so-called
gamma
metric ($\gamma $-metric) \cite{9}, \cite{10}. This metric, which is
also
known as Zipoy-Vorhees metric \cite{11}, belongs to the family of Weyl's
solutions, and is continuously linked to the Schwarzschild space-time
through one of its parameters. The motivation for this choice is
twofold. On
one hand the exterior $\gamma $-metric corresponds to a solution of the
Laplace equation (in cylindrical coordinates) with the same singularity
structure as the Schwarzschild solution (a line segment \cite{9}). In
this
sense the $\gamma $-metric appears as the ``natural'' generalization of
Schwarzschild space-time to the axisymmetric case. On the other hand,
the
two interior solutions considered have reasonable physical properties
and
generalize important and useful sources of the Schwarzschild space-time,
namely the interior Schwarzschild solution (homogeneous density) and the
Adler solution \cite{12}.

From all the comments above, it is not difficult to infer the astrophysical
relevance of the results here presented. Indeed,
spherical symmetry is a common assumption in the study of compact
self-gravitating objects (white dwarfs, neutron stars, black holes
) , furthermore in the specific case of non-rotating black holes, spherical
symmetry should be "absolute", according to Israel
theorem. Therefore it is pertinent to ask, how do small deviations from
this assumption, related to any kind of perturbation
(e.g. fluctuations of the stellar matter, external perturbations, etc),
affect the dynamics of the system?. The result obtained here,
 which deserves to be  to emphasized, is that slight deviations from
spherical symmetry seriously modify the departure from
equilibrium in the two examples presented. On the other hand, we are well
aware of the fact that the $\gamma $ metric is not the only
possible description for the exterior of a compact objetc and, of course,
the two equations of state considered here do not exhaust
the list of possible candidates for the equation of state of the stellar
matter.However, in view of the properties of the $\gamma $
metric and the two equations of state considered here, mentioned above, it
is fair to say that the case for the relevance of small
deviations from spherical symmetry in the dynamics of compact objects has
been established. This means that any conclusion on the
structure and evolution of a compact object, derived on the assumption of
spherical symmetry should be carefully   checked against
deviations from that assumption.

This paper is organized as follows. In the next section we shall specify
the
space-time inside and outside the matter distribution, and the
conventions
used. In section 3 we give the
energy momentum tensor components in terms of variables measured by a
locally Minkowskian and comoving observer. The departure from
equilibrium is analyzed in section 4, and results are discussed in the
last
section.
We have included an appendix containing the components of the Einstein tensor,
the Einstein equations in the limit of small non sphericity and the
components of the
conservation law.

\section{The Space-time}

\subsection{The exterior space-time}

As has been mentioned above, our initial matter configuration is the
source
of an axially symmetric and static space-time ($\gamma $-metric).
In cylindrical
coordinates,
static axisymmetric solutions to Einstein equations are given by the
Weyl
metric \cite{2}
\begin{equation}
ds^2=e^{2\lambda }dt^2-e^{-2\lambda }\left[ e^{2\mu }\left( d\rho
^2+dz^2\right) +\rho ^2d\varphi ^2\right] ,  \label{1}
\end{equation}
with
\begin{equation}
\lambda _{,\rho \rho }+\rho ^{-1}\lambda _{,\rho }+\lambda _{,zz}=0
\label{2}
\end{equation}
and
\begin{equation}
\mu _{,\rho }=\rho \left(\lambda _{,\rho }^2-\lambda _{,z}^2\right)\qquad
\mu _{,z}=2\rho
\lambda _{,\rho }\lambda _{,z}.  \label{3}
\end{equation}
Observe that (\ref{2}) is just the Laplace
equation for $\lambda $ (in the Euclidean space).

The $\gamma $-metric is defined by \cite{9}
\begin{equation}
\lambda =\frac \gamma 2\ln \left[ \frac{R_1+R_2-2m}{R_1+R_2+2m}\right] ,
\label{4}
\end{equation}
\begin{equation}
e^{2\mu} =\left[ \frac{\left( R_1+R_2+2m\right) \left( R_1+R_2-2m\right)
}{%
4R_1R_2}\right] ^{\gamma ^2},  \label{5}
\end{equation}
where
\begin{equation}
R_1^2=\rho ^2+(z-m)^2\qquad R_2^2=\rho ^2+(z+m)^2.  \label{6}
\end{equation}
It is worth noticing that $\lambda ,$ as given by (\ref{4}), corresponds
to
the Newtonian potential of a line segment of mass density $\gamma/2 $ and
length $2m,$ symmetrically distributed along the $z$ axis. The
particular
case $\gamma =1,$ corresponds to the Schwarzschild metric.

It will be useful to work in Erez-Rosen coordinates \cite{11}, given by
\begin{equation}
\rho ^2=(r^2-2mr)\sin ^2\theta \qquad z=(r-m)\cos \theta ,  \label{7}
\end{equation}
which yields the line element as \cite{9}
\begin{equation}
ds^2=Fdt^2-F^{-1}\left\{ Gdr^2+Hd\theta ^2+\left( r^2-2mr\right) \sin
^2\theta d\varphi ^2\right\} ,  \label{8}
\end{equation}
where
\begin{eqnarray}
F &=&\left( 1-\frac{2m}r\right) ^\gamma ,  \label{8a} \\
&&  \nonumber \\
G &=&\left( \frac{r^2-2mr}{r^2-2mr+m^2\sin ^2\theta }\right) ^{\gamma
^2-1},
\label{8b}
\end{eqnarray}
and
\begin{equation}
H=\frac{\left( r^2-2mr\right) ^{\gamma ^2}}{\left( r^2-2mr+m^2\sin
^2\theta
\right) ^{\gamma ^2-1}}  \label{8c}
\end{equation}
Now, it is easy to check that $\gamma =1$ corresponds to the
Schwarzschild
metric.

The total mass of the source is \cite{9,10} $M=\gamma m,$ and the
quadrupole
moment is given by
\begin{equation}
Q=\frac \gamma 3m^3\left( 1-\gamma ^2\right) .  \label{9}
\end{equation}
So that $\gamma >1$ ($\gamma <1$) corresponds to an oblate (prolate)
spheroid.

\subsection{The interior space-time}

The metric within the matter distribution bounded by the surface
\begin{equation}
r=a  \label{10}
\end{equation}
is given by
\begin{eqnarray}
g_{tt} &=&f^{2\gamma }  \nonumber \\
g_{rr} &=&-f^{2(1-\gamma )}\Delta ^{\gamma ^2-2}\Sigma ^{1-\gamma ^2}
\nonumber \\
g_{\theta \theta } &=&-r^2f^{2\gamma (\gamma -1)}\Phi ^{1-\gamma^2}
\nonumber \\
g_{\varphi \varphi } &=&-r^2f^{2(1-\gamma )}\sin ^2\theta
\label{11}
\end{eqnarray}
where $f,$ $\Delta ,$ $\Sigma $ and $\Phi $ are functions whose specific
form depends on the model under consideration.

The two cases to be considered here are the solutions reported in
\cite{7},
namely

\begin{enumerate}
\item  The modified constant density Schwarzschild solution
\begin{eqnarray}
f(r)&=&\frac 32\sqrt{1-\frac{a^2}{B^2}}-\frac 12\sqrt{1-\frac{r^2}{B^2}}
\label{12} \\
\nonumber \\
\Delta (r)&=&1-\frac{r^2}{B^2}  \label{13} \\
\nonumber \\
\Sigma (r,\theta )&=&1-\frac{r^2}{B^2}+\frac{r^4}{4B^4}\sin ^2\theta
\label{14} \\
\nonumber \\
\Phi (r,\theta )&=&f^2+\frac{r^4}{4B^4}V(r)\sin ^2\theta   \label{15}
\end{eqnarray}
with
\begin{equation}
V(r)=1+\frac 6a\left( a-r\right) ,  \label{16}
\end{equation}
and
\begin{equation}
B^2=\frac 3{8\pi \rho _{ss}},  \label{17}
\end{equation}
where $\rho _{ss}$ denotes the energy density in the spherically
symmetric
limit ($\gamma =1$)

\item  The modified Adler solution
\begin{eqnarray}
f(r)&=&A+Br^2  \label{18} \\
\nonumber \\
\Delta (r)&=&1+\frac{Cr^2}{\left( A+3Br^2\right) ^{2/3}}  \label{18a} \\
\nonumber \\
\Sigma (r,\theta )&=&1+\frac{Cr^2}{\left( A+3Br^2\right)
^{2/3}}+\frac{C^2r^4}{%
4\left( A+3Br^2\right) ^{4/3}}\sin ^2\theta   \label{19} \\
\nonumber \\
\Phi (r,\theta )&=&\left( A+Br^2\right) ^2+\frac{C^2r^4V(r)}{4\left(
A+3Br^2\right) ^{4/3}}\sin ^2\theta   \label{20} \\
\end{eqnarray}
with
\begin{equation}
V(r)=1+\frac 6a\left( 1-\frac{5m}{3a}\right) \left( 1-\frac ma\right)
^{-1}\left( a-r\right)   \label{21}
\end{equation}
and
\begin{eqnarray}
A&=&\frac{1-\frac{5m}{2a}}{\left(1-\frac{2m}{a}\right)^{1/2}}
\nonumber \\
B&=&\frac m{2a^3\left( 1-\frac{2m}a\right) ^{1/2}}  \nonumber \\
\nonumber \\
C&=&-\frac{2m\left( 1-\frac ma\right) ^{2/3}}{a^3\left( 1-\frac{2m}{a}\right)
^{1/3}}  \label{22}
\end{eqnarray}
\end{enumerate}

Before closing this section, two remarks are in order:

\begin{enumerate}
\item  Since we are considering the source described in (\ref{11}) as an
initial state, the time derivatives of functions $f,$ $\Delta ,$ $\Sigma$
and $\Phi $ will be in principle different from zero.

\item  Junction (Darmois) conditions are satisfied at the boundary $r=a$
-see \cite{7} for details.
\end{enumerate}

\section{The energy momentum tensor}

In order to give physical meaning to the components of the energy
momentum
tensor in coordinates ($t,$ $r,$ $\theta ,$ $\varphi $), we shall
develop a
procedure similar to that used by Bondi \cite{13} in his study of non
static
spherically symmetric sources. Thus, we introduce purely local Minkowski
coordinates ($\tau ,$ $x,$ $y,$ $z$) defined by
\begin{eqnarray}
d\tau &=&f^\gamma dt  \label{31} \\
&&  \nonumber \\
dx &=&f^{1-\gamma }\Delta ^{-1+\gamma ^2/2}\Sigma ^{(1-\gamma ^2)/2}dr
\label{32} \\
&&  \nonumber \\
dy &=&rf^{\gamma (\gamma -1)}\Phi ^{(1-\gamma ^2)/2}d\theta  \label{33}
\\
&&  \nonumber \\
dz &=&r\sin (\theta )f^{1-\gamma }d\varphi .  \label{34}
\end{eqnarray}

Next, since we are assuming that our source does not dissipate energy,
then
the covariant components of the energy momentum tensor, as measured by a
local Minkowskian and comoving with the fluid observer, will be
\begin{equation}
\widehat{T}_{\mu \nu }=\left(
\begin{array}{cccc}
\rho & 0 & 0 & 0 \\
0 & p_{xx} & p_{xy} & 0 \\
0 & p_{yx} & p_{yy} & 0 \\
0 & 0 & 0 & p_{zz}
\end{array}
\right) ,  \label{35}
\end{equation}
where $\rho $ is the energy density and in general $p_{xx}\neq
p_{yy}\neq
p_{zz}$ and $p_{xy}=p_{yx}.$ We may write (\ref{35}) in the form
\begin{equation}
\widehat{T}_{\mu \nu }=\left( \rho +p_{zz}\right) \widehat{U}_\mu
\widehat{U}%
_\nu -p_{zz}\eta _{\mu \nu }+\left( p_{xx}-p_{zz}\right) \widehat{k}_\mu
\widehat{k}_\nu +\left( p_{yy}-p_{zz}\right) \widehat{l}_\mu
\widehat{l}_\nu
+2p_{xy}\widehat{k}_{(\mu }\widehat{l}_{\nu )},  \label{36}
\end{equation}
where $\eta _{\mu \nu }$ denotes the flat space-time metric and
\begin{eqnarray}
\widehat{U}_\mu &=&\left(
\begin{array}{cccc}
1, & 0, & 0, & 0
\end{array}
\right)  \label{37} \\
&&  \nonumber \\
\widehat{k}_\mu &=&\left(
\begin{array}{cccc}
0, & 1, & 0, & 0
\end{array}
\right)  \label{38} \\
&&  \nonumber \\
\widehat{l}_\mu &=&\left(
\begin{array}{cccc}
0, & 0, & 1, & 0
\end{array}
\right)  \label{39b}
\end{eqnarray}

The components of the energy-momentum tensor $T_{\mu \nu }$ in ($t,$
$r,$ $%
\theta ,$ $\varphi $) coordinates are linked to (\ref{36}) by
\begin{equation}
T_{\mu \nu }=\Lambda _\mu ^\alpha \Lambda _\nu ^\beta L_\alpha ^\gamma
L_\beta ^\delta \widehat{T}_{\gamma \delta },  \label{40}
\end{equation}
where $\Lambda _\mu ^\nu =\partial x^\nu /\partial x^\mu $ is the local
coordinate transformation matrix between Minkowskian coordinates and
($t,$ $%
r,$ $\theta ,$ $\varphi $) coordinates and the Lorentz matrices $L_\mu ^\nu
$
are given by
\begin{equation}
L_t^t=\Gamma \qquad L_i^t=L_t^i=-\Gamma w_i\qquad L_j^i=L_i^j=\delta
_j^i+%
\frac{(\Gamma -1)w_iw_j}{w^2},  \label{41}
\end{equation}
where
\begin{equation}
w^2=w_x^2+w_y^2\qquad \Gamma =\frac 1{\sqrt{1-w^2}},  \label{42}
\end{equation}
and $w_x$ and $w_y$ denote, respectively, the velocity of a fluid
element
along the $x$  and $y$ ($r$ and $\theta $) directions, as measured by our
local
Minkowskian observer as defined by (\ref{31})--(\ref{34}). Observe that we
are
considering the case $w_z=0,$ which means that the system preserves the
reflection symmetry (besides the axial symmetry).

The non vanishing components of $\Lambda _\mu ^\nu $ are
\begin{eqnarray}
\Lambda _t^\tau &=&f^\gamma  \label{43} \\
&&  \nonumber \\
\Lambda _r^x &=&f^{1-\gamma }\Delta ^{-1+\gamma ^2/2}\Sigma ^{(1-\gamma
^2)/2}  \label{44} \\
&&  \nonumber \\
\Lambda _\theta ^y &=&rf^{\gamma (\gamma -1)}\Phi ^{(1-\gamma ^2)/2}
\label{45} \\
&&  \nonumber \\
\Lambda _\varphi ^z &=&r\sin (\theta )f^{1-\gamma }.  \label{46}
\end{eqnarray}

Then, (\ref{40}) readly gives
\begin{eqnarray}
T_{tt} &=&f^{2\gamma }\Gamma ^2\left( \rho
+p_{xx}w_x^2+p_{yy}w_y^2+2p_{xy}w_xw_y\right)   \label{47} \\
&&  \nonumber \\
T_{tr} &=&-f\Delta ^{-1+\gamma ^2/2}\Sigma ^{(1-\gamma ^2)/2}\Gamma
\times
\nonumber \\
&&\left( \Gamma w_x\rho +p_{xx}w_x\Lambda _x+p_{yy}w_y\Lambda
+p_{xy}\left[
w_x\Lambda +w_y\Lambda _x\right] \right)   \label{48} \\
&&  \nonumber \\
T_{t\theta } &=&-rf^{\gamma ^2}\Phi ^{(1-\gamma ^2)/2}\Gamma \times
\nonumber \\
&&\left( \Gamma w_y\rho +p_{xx}w_x\Lambda +p_{yy}w_y\Lambda
_y+p_{xy}\left[
w_y\Lambda +w_x\Lambda _y\right] \right)   \label{49} \\
&&  \nonumber \\
T_{rr} &=&f^{2-2\gamma }\Delta ^{\gamma ^2-2}\Sigma ^{1-\gamma ^2}\times
\nonumber \\
&&\left( \Gamma ^2w_x^2\rho +p_{xx}\Lambda _x^2+p_{yy}\Lambda
^2+2p_{xy}\Lambda \Lambda _x\right)   \label{50} \\
&&  \nonumber \\
T_{r\theta } &=&rf^{(\gamma -1)^2}\Delta ^{-1+\gamma ^2/2}\Phi
^{(1-\gamma
^2)/2}\Sigma ^{(1-\gamma ^2)/2}\times  \nonumber \\
&&\left( \Gamma ^2w_xw_y\rho +\Lambda \left[ p_{xx}\Lambda
_x+p_{yy}\Lambda
_y\right] +p_{xy}\left[ \Lambda ^2+\Lambda _x\Lambda _y\right] \right)
\label{51} \\
&&  \nonumber \\
T_{\theta \theta } &=&r^2f^{2\gamma (\gamma -1)}\Phi ^{1-\gamma
^2}\times
\nonumber \\
&&\left( \Gamma ^2w_y^2\rho +p_{xx}\Lambda ^2+p_{yy}\Lambda
_y^2+2p_{xy}\Lambda \Lambda _y\right)   \label{52} \\
&&  \nonumber \\
T_{\varphi \varphi } &=&r^2f^{2(1-\gamma )}\sin ^2(\theta )p_{zz},
\label{53}
\end{eqnarray}
with
\begin{eqnarray}
\Lambda  &\equiv &\frac{\left( \Gamma -1\right) w_xw_y}{w^2},
\label{53a} \\
&&  \nonumber \\
\Lambda _x &\equiv &1+\frac{\left( \Gamma -1\right) w_x^2}{w^2},
\label{53b}
\\
&&  \nonumber \\
\Lambda _y &\equiv &1+\frac{\left( \Gamma -1\right) w_y^2}{w^2},
\label{53c}
\end{eqnarray}

So that
\begin{equation}
{T}_{\mu \nu }=\left( \rho +p_{zz}\right) {U}_\mu
{U}%
_\nu -p_{zz} g_{\mu \nu }+\left( p_{xx}-p_{zz}\right) {k}_\mu
{k}_\nu +\left( p_{yy}-p_{zz}\right) {l}_\mu
{l}_\nu
+2p_{xy}{k}_{(\mu }{l}_{\nu )},  \label{Tmn}
\end{equation}
where $U_\mu$, $k_\mu$ and $l_\mu$ are obtained
after applying the boost velocity (\ref{41}) and the coordinate
transformation (\ref{43})--(\ref{46}) to (\ref{37})--(\ref{39b}),
\begin{equation}
U_\mu =\Gamma \left(
\begin{array}{cccc}
f^\gamma , & -w_xf^{1-\gamma }\Delta ^{-1+\gamma ^2/2}\Sigma ^{(1-\gamma
^2)/2}, & -w_yrf^{\gamma (\gamma -1)}\Phi ^{(1-\gamma ^2)/2}, & 0
\end{array}
\right) ,  \label{53e}
\end{equation}
\begin{equation}
k_\mu =\left(
\begin{array}{cccc}
-\Gamma w_xf^\gamma , & f^{1-\gamma }\Delta ^{-1+\gamma ^2/2}\Sigma
^{(1-\gamma ^2)/2}\Lambda _x, & rf^{\gamma (\gamma -1)}\Phi ^{(1-\gamma
^2)/2}\Lambda , & 0
\end{array}
\right) ,  \label{53f}
\end{equation}
\begin{equation}
l_\mu =\left(
\begin{array}{cccc}
-\Gamma w_yf^\gamma , & f^{1-\gamma }\Delta ^{-1+\gamma ^2/2}\Sigma
^{(1-\gamma ^2)/2}\Lambda , & rf^{\gamma (\gamma -1)}\Phi ^{(1-\gamma
^2)/2}\Lambda _y, & 0
\end{array}
\right) .  \label{53g}
\end{equation}

\section{Departure from equilibrium}

Let us consider a static axially symmetric source defined by (\ref{11}),
which once submitted to perturbations, departs from equilibrium without
dissipation. We shall evaluate the system after such departure, on a
time
scale such that $w_x$ and $w_y$ remain vanishingly small, whereas their
time
derivatives though small, will be different from zero.

Thus, just after leaving the equilibrium, the following conditions hold
\begin{equation}
w_x=w_y=w_{x,i}=w_{y,i}\simeq 0,\qquad (i=r,\theta ,\varphi )
\label{58}
\end{equation}
\begin{equation}
w_{x,t},\ w_{y,t}\neq 0\qquad (\mbox{small})  \label{59}
\end{equation}

From now on, unless otherwise stated, all equations are evaluated
at the moment the system starts to deviate from equilibrium.

Then from (\ref{47})--(\ref{53}), we obtain using (\ref{58})

\begin{equation}
T_{t \theta} = T_{t r} = 0
\label{T0}
\end{equation}
which implies, because of (\ref{26}) and (\ref{28})
\begin{equation}
\Delta_{,t} = f_{,t} = \Sigma_{,t} = \Phi_{,t} = 0
\label{der0}
\end{equation}
where for simplicity we write $0$ for ${\cal O}(\omega)$ (as we shall do
hereafter).

Obviously, spatial derivatives of the above quantities will be also
vanishingly small on the time scale under consideration.

Next, we shall evaluate the conservation law $T_{\nu;\mu}^{\mu}=0$.

The $t$-component yields
\begin{equation}
\rho_{,t}=0  \label{62}
\end{equation}

whereas the $r$ and $\theta$ component lead, after inspection of (\ref{A2})
and (\ref{A3}), to

\begin{eqnarray}
& - & f^{-1} f^{-2\varepsilon} \Delta^{-1/2} \Delta^{(\varepsilon +
\varepsilon^2/2)}
\Sigma^{-(\varepsilon + \varepsilon^2/2)} \{\omega_{x,t}(\rho + p_{xx}) +
\omega_{y,t} p_{xy}\}\nonumber \\
& - & \left[ \frac{f_{,r}}{f} (1 + \varepsilon + \varepsilon^2) +
\frac{\Phi_{,r}}{2\Phi} (-2\varepsilon - \varepsilon^2)
+ \frac{2}{r}\right] p_{xx} - p_{xx,r}\nonumber \\
& - & r^{-1} f^{-(2\varepsilon + \varepsilon^2)} \Delta^{-1/2}
\Delta^{(\varepsilon + \varepsilon^2/2)}
\Phi^{(\varepsilon + \varepsilon^2/2)} \Sigma^{-(\varepsilon +
\varepsilon^2/2)} \nonumber \\
& \times & \left\{\left[\frac{\Sigma_{,\theta}}{\Sigma}(-2\varepsilon -
\varepsilon^2) + \cot{\theta}\right] p_{xy}
+ p_{xy,\theta}\right\}\nonumber \\
& - & \frac{f_{,r}}{f} (1+\varepsilon) \rho + \left[\frac{1}{r} +
\frac{f_{,r}}{f} (-\varepsilon)\right] p_{zz} \nonumber \\
& + & \left[\frac{1}{r}+ \frac{f_{,r}}{f}(\varepsilon+\varepsilon^2) +
\frac{\Phi_{,r}}{2\Phi}(-2\varepsilon-\varepsilon^2)\right] p_{yy} = 0
\label{uno}
\end{eqnarray}
and
\begin{eqnarray}
& - & r f^{-1} f^{\varepsilon^2} \Phi^{(-\varepsilon-\varepsilon^2/2)}
\{\omega_{y,t}(\rho + p_{yy}) + \omega_{x,t} p_{xy}\}\nonumber \\
& - & r f^{(2\varepsilon + \varepsilon^2)} \Delta^{1/2}
\Delta^{-(\varepsilon + \varepsilon^2/2)}
\Phi^{-(\varepsilon + \varepsilon^2/2)} \Sigma^{(\varepsilon +
\varepsilon^2/2)} \nonumber \\
& \times & \left\{\left[\frac{f_{,r}}{f}(1+2\varepsilon+2\varepsilon^2) +
\frac{\Phi_{,r}}{\Phi}
(-2\varepsilon - \varepsilon^2) + \frac{3}{r}\right]p_{xy} +
p_{xy,r}\right\}\nonumber \\
& - & \left[\frac{\Sigma_{,\theta}}{2\Sigma} (-2\varepsilon -
\varepsilon^2) + \cot{\theta}\right] p_{yy}\nonumber \\
& + & \cot{\theta} p_{zz} - p_{yy,\theta} +
\frac{\Sigma_{,\theta}}{\Sigma}(-2\varepsilon - \varepsilon^2) p_{xx}= 0
\label{dos}
\end{eqnarray}
where we have assumed
$$
\gamma = 1 + \varepsilon
$$
and $\varepsilon$ may be either positive or negative.

Before going further into the analysis of the equations above it is quite
instructive
to consider the spherically symmetric situation. In this case we have

\begin{equation}
\Sigma_{,\theta}= \omega_{y,t} = p_{xy} = p_{zz}-p_{yy} = p_{yy,\theta} =
\varepsilon = 0
\label{tre}
\end{equation}

Then, (\ref{dos}) becomes an identity and (\ref{uno}) reads

\begin{equation}
- \omega_{x,t} (\rho + p_{xx}) = \left[p_{xx,r}+\frac{2}{r}(p_{xx}-p_{yy})+
\frac{\nu_{,r}}{2}(\rho + p_{xx})\right]
e^{(\nu-\delta)/2}
\label{cua}
\end{equation}
where
\begin{equation}
\nu_{,r}= 2 \frac{m + 4\pi r^3 p_{xx}}{r(r-2m)}
\label{cin}
\end{equation}
\begin{equation}
f=e^{\nu/2} \qquad ; \qquad \Delta \equiv e^{-\delta} = 1 - \frac{2m}{r}
\label{sei}
\end{equation}
with
\begin{equation}
m = \int^r_0{4\pi r^2 \rho dr}
\label{sie}
\end{equation}

The physical meaning of (\ref{cua}) is quite transparent. It has the
``Newtonian'' form
$$
Force = mass \times acceleration
$$
Indeed, the left-hand side consists of two factors. The time derivative of
radial
velocity and the inertial mass density. On the right-hand side,
we have three possible sources of forces:
The pressure gradient (negative), the possible
anisotropic pressure contribution and the ``gravitational''
term (positive), multiplied by the relativistic correction
factor $e^{(\nu-\delta)/2}$. As the source becomes more and more
compact the relativistic correction factor decreases as $\approx r-2m$,
whereas the ``gravitational'' term grows as $\approx 1/(r-2m)$.
Since this latter term is positive and will prevail (as $r \rightarrow 2m$)
over
the two other force terms, we are lead unavoidably to a catastrophic collapse
($\omega_{x,t}<0$), independently on the equation of state of the configuration.

Let us now turn back to (\ref{uno}) and (\ref{dos}), to infer what happens
in the
non-spherical case (i.e. $\varepsilon \not= 0$), even though $\varepsilon<<1.$
Then neglecting higher order terms on $\varepsilon$, and being careful with
metric functions terms (some of which may tend to zero as the object
becomes more and more compact)
we obtain from (\ref{uno})
\begin{eqnarray}
& - & f^{-2\varepsilon}\left[\omega_{x,t}(\rho+p_{xx})\right]=
f \Delta^{1/2}\Delta^{-\varepsilon} \Sigma^{\varepsilon}
\left(\frac{f_{,r}}{f}\right)(\rho+p_{xx})(1+\varepsilon)
\nonumber \\
& + & f \Delta^{1/2} \Delta^{-\varepsilon} \Sigma^{\varepsilon}
\left\{p_{xx,r}- \frac{1}{r}
\left[-2p_{xx}+p_{yy}+p_{zz}\right] - \varepsilon
\frac{\Phi_{,r}}{\Phi}(p_{xx} - p_{yy})\right\}\nonumber \\
& - & r^{-1} f^{1-2\varepsilon} \Phi^{\varepsilon}\left[\cot{\theta}
p_{xy}+p_{xy,\theta}\right]
\label{och}
\end{eqnarray}
where we have used the fact that
$$
\omega_{y,t} \approx p_{xy} \approx p_{zz} - p_{yy} \approx O(\varepsilon)
$$
Since in this approximation the system (\ref{uno}), (\ref{dos}) is not
longer coupled (in $\omega_{x,t}$ and $\omega_{y,t}$)
we shall consider only equation (\ref{och}).

In order to extract more information from (\ref{och}) it is necessary to
specify the source under consideration.
We shall use the two configurations mentioned in section 2.
In both cases, $f$ vanishes before $a=2m$. Thus, in the Schwarzschild-like
models we have $f(0)=0$, if $2m/a = 8/9$.
Since we know that any spherically symmetric static configuration with
constant $\rho_{ss}$ and locally
isotropic pressure should satisfy the constraint $n \equiv 2M/a<8/9$ we may
assume that the system leaves the equilibrium
for values of $n$ close to $8/9$.
Then, an inspection of (\ref{och}) shows that (if $\varepsilon>0$), for
values of $n$ approaching
$8/9$, the values of $\omega_{x,t}$ should become
vanishingly small.
Indeed, as $n \rightarrow 8/9$, $f \rightarrow 0$ and $p_{xx} \approx 1/f$
; $f_{,r}/f \approx p_{xx}$,
whereas $\Delta$ and $\Sigma$ remain different from zero and bounded.
Then, the largest term on the right of (\ref{och}) will be of
order $p_{xx}$ (or $1/f$).
So, for any finite $\varepsilon$ (no matter how small) $\omega_{x,t}$
should be of order $f^{2\varepsilon}$.

We may now use Einstein equations to
elucidate that for negative values of $\varepsilon$ the system may be
unphysical.
If $n$ takes values close to the limit allowed by the model
(8/9 for Schwarzschild-type model and 4/5 for Adler-type model),
the critical values ($f\rightarrow 0$, $p_{xx}\rightarrow 1/f$ ...)
appear close to $r=0$. Outside of this region the system is basically
composed by an spherical incompressible fluid plus a perturbation in
$\varepsilon$.
Thus, the physical or unphysical character of the model is determined
by its behaviour close to these two limits.

The energy density, in the limit $r\rightarrow 0$,  for
the Schwarzschild-type model and Adler-type model is given
by expresions (\ref{schdensid}) and (\ref{adldensid}) respectively.
From these ones, it is easy to show that if $\varepsilon<0$,
the energy density becomes negative
as the system approaches to $n\rightarrow 8/9$ (Schwarzschild case)
or $n\rightarrow 4/5$ (Adler case) and positive for $\varepsilon \ge 0$.
Therefore, in both cases, positive energy conditions impose $\varepsilon \ge 0$,
and for very compact objects (close to the limit allowed by the model)
the inertial mass term substantially increases suggesting the "stalling" of the
collapse, close to, but before, the maximum allowed value of $n$.

\section{Conclusions}

We have seen so far that (as expected), for high gravitational fields,
important differences appear between the spherical and the non-spherical
collapse. This conclusion being true even for small non-sphericity.

In the two models considered above, a factor multiplying the inertial
mass
term bring out those differences. Alternatively we may multiply both sides
of equation (\ref{och}) by $f^{2\varepsilon}$ and say that the total "force
term"
decreases as $f^{2\varepsilon}$. The final result being the same,
namely that $\omega_{x,t}\approx f^{2\varepsilon}$. Although for the two
examples
considered
here, we have considered $\varepsilon >0,$ it is obvious that
there
exist models with $\varepsilon <0,$ in which case the effective inertial
mass density term may become very small, leading to highly unstable
situations and to a breakdown of the linear approximation.
In fact, it is worth noticing that important differences between the two
cases ($\varepsilon >0$ and $\varepsilon <0$) appear also in the
behaviour of the exterior $\gamma $-metric (\cite{9}, \cite{10})

Thus, on the basis of presented results we may conclude that whatever
the
model and the sign of $\varepsilon $ would be, any source of Weyl metric
would evolve quite differently from the corresponding spherical source, as
critical values of $n$ are considered.

In particular, in the example examined, the inevitability of collapse in
the
spherical case, appears to be modified by a sharp increase in the
effective
inertial mass density term (or a sharp decrease in the "total force" term)
 as $n$ approaches its maximum allowed value.
This
increase makes the system more stable, hindering its departure from
equilibrium.

We would like to conclude by stressing our main point: the inevitability
of the catastrophic collapse for very compact objects, which appears in
the spherically symmetric case, (for any regular matter configuration) is
not present at least for the family of solutions considered here (e.g. is
not ``strictly'' unavoidable).

\section*{Acknowledgment}
We are deeply indebted to Professor Bondi for his comments and criticisms
and his
generous and encouraging support.
One of us (J.M.) would like to express his thanks to the Theoretical Physics
Group for hospitality at the Physics Department of the University of Salamanca.
This work was partially supported by the Spanish Ministry of Education
under Grant No. PB94-0718.
L.H. wishes to thank Lou Witten for interesting comments.

\appendix

\section{The Einstein tensor}

The calculation of non-vanishing components of the Einstein tensor
\begin{equation}
G_{\mu \nu }\equiv R_{\mu \nu }-\frac 12g_{\mu \nu }R  \label{23}
\end{equation}
yields for the metric (\ref{11}), using Maple V with GrTensor library
and
checking against the results given in \cite{7bis},
\begin{eqnarray}
G_{rr} &=&-\Sigma ^{1-\gamma ^2}\Delta ^{\gamma ^2-2}f^{2-4\gamma
}\times
\nonumber \\
&&\{  \nonumber \\
&&\frac{f_{,tt}}f(1-\gamma )^2+\frac{\Phi _{,tt}}{2\Phi }(1-\gamma
^2)+\left( \frac{\Phi _{,t}}{2\Phi }\right) ^2(\gamma ^4-1)+\left(
\frac{%
f_{,t}}f\right) ^2\gamma (\gamma -2)(\gamma -1)^2  \nonumber \\
&&  \nonumber \\
&&+\frac{\Phi _{,t}}{2\Phi }\frac{f_{,t}}f(1-\gamma ^2)(2\gamma
^2-4\gamma
+1)  \nonumber \\
&&\}  \nonumber \\
&&  \nonumber \\
&&-r^{-2}\Sigma ^{1-\gamma ^2}\Delta ^{\gamma ^2-2}f^{2-2\gamma ^2}\Phi
^{\gamma ^2-1}\left[ 1+\frac{\Phi _{,\theta }}{2\Phi }(1-\gamma ^2)\cot
\theta \right]  \nonumber \\
&&  \nonumber \\
&&+\frac{\Phi _{,r}}{2\Phi }(1-\gamma ^2)\left[ \frac{f_{,r}}f+\frac
1r\right] +\frac{f_{,r}}f\frac 1r(1+\gamma ^2)+\frac 1{r^2}  \label{24}
\end{eqnarray}

\begin{equation}
G_{r\theta }=(1-\gamma ^2)\left[ \frac{\Sigma _{,\theta }}{2\Sigma
}\left(
\frac{f_{,r}}f+\frac 1r\right) +\left( \frac{\Phi _{,r}}{2\Phi
}-\frac{f_{,r}%
}f\right) \cot \theta \right]  \label{25}
\end{equation}

\begin{eqnarray}
G_{rt} &=&-\frac{f_{,rt}}f(\gamma -1)^2+\frac{\Phi _{,rt}}{2\Phi
}(\gamma
^2-1)  \nonumber \\
&&  \nonumber \\
&&+\left( \frac{\Delta _{,t}}{2\Delta }(\gamma ^2-2)+(1-\gamma ^2)\left[
\frac{\Sigma _{,t}}{2\Sigma }+\frac{f_{,t}}f\right] \right) \left[
\frac{%
f_{,r}}f(1-\gamma )^2+\frac{\Phi _{,r}}{2\Phi }(1-\gamma ^2)\right]
\nonumber \\
&&  \nonumber \\
&&+\frac 1r\left( \frac{\Delta _{,t}}\Delta (\gamma ^2-2)+(1-\gamma
^2)\left[ \frac{\Sigma _{,t}}\Sigma +\frac{f_{,t}}f\right] \right)
\nonumber
\\
&&  \nonumber \\
&&+\frac{\Phi _{,t}}{2\Phi }\left[ \frac{f_{,r}}f\gamma \left( \gamma
^2-1\right) (\gamma -2)+\frac{\Phi _{,r}}{2\Phi }(1-\gamma
^4)+\frac{(\gamma
^2-1)}r\right]  \label{26}
\end{eqnarray}

\begin{eqnarray}
G_{\theta \theta } &=&r^2f^{2\gamma ^2-2}\Sigma ^{\gamma ^2-1}\Delta
^{2-\gamma ^2}\Phi ^{1-\gamma ^2}\times  \nonumber \\
&&\{  \nonumber \\
&&\frac{f_{,rr}}f+\left( \frac 1r+\frac{f_{,r}}f\right) \left[
\frac{\Sigma
_{,r}}{2\Sigma }(\gamma ^2-1)-\frac{\Delta _{,r}}{2\Delta }(\gamma
^2-2)\right] +\frac{f_{,r}}f\left( \frac 1r+(\gamma ^2-1)\frac{f_{,r}}%
f\right)  \nonumber \\
&&\}  \nonumber \\
&&  \nonumber \\
&&+r^2f^{2\gamma (\gamma -2)}\Phi ^{1-\gamma ^2}\times  \nonumber \\
&&\{  \nonumber \\
&&\frac{\Delta _{,t}}{2\Delta }(\gamma ^2-2)\left[ \frac{\Delta _{,t}}{%
2\Delta }(4-\gamma ^2)+\frac{\Sigma _{,t}}\Sigma (\gamma
^2-1)+\frac{f_{,t}}%
f(4\gamma -3)\right]  \nonumber \\
&&  \nonumber \\
&&+\frac{\Sigma _{,t}}{2\Sigma }(1-\gamma ^2)\left[ \frac{\Sigma
_{,t}}{%
2\Sigma }(1+\gamma ^2)+\frac{f_{,t}}f(4\gamma -3)\right] +\left(
\frac{f_{,t}%
}f\right) ^2(1-\gamma )(5\gamma -1)  \nonumber \\
&&  \nonumber \\
&&+\frac{\Delta _{,tt}}{2\Delta }(2-\gamma ^2)+\frac{\Sigma
_{,tt}}{2\Sigma }%
(\gamma ^2-1)+\frac{2f_{,tt}}f(\gamma -1)  \nonumber \\
&&\}  \nonumber \\
&&  \nonumber \\
&&+\frac{\Sigma _{,\theta }}{2\Sigma }(1-\gamma ^2)\cot \theta
\label{27}
\end{eqnarray}

\begin{eqnarray}
G_{\theta t} &=&\left( \frac{\Phi _{,t}}{2\Phi }-\frac{f_{,t}}f\right)
(1-\gamma ^2)\cot \theta +\frac{\Sigma _{,t\theta }}{2\Sigma }(\gamma
^2-1)-%
\frac{\Sigma _{,\theta }}{2\Sigma }(1-\gamma ^2)\times  \nonumber \\
&&  \nonumber \\
&&\left[ \frac{f_{,t}}f(1-\gamma ^2)+\frac{\Delta _{,t}}{2\Delta
}(\gamma
^2-2)-\frac{\Phi _{,t}}{2\Phi }(1-\gamma ^2)-\frac{\Sigma _{,t}}{2\Sigma
}%
(1+\gamma ^2)\right]  \label{28}
\end{eqnarray}

\begin{eqnarray}
G_{\varphi \varphi } &=&r^2\sin ^2\theta \Delta ^{2-\gamma ^2}\Sigma
^{\gamma ^2-1}\times  \nonumber \\
&&\{  \nonumber \\
&&\frac{\Phi _{,r}}{2\Phi }\left[ (\gamma ^2-1)\left( \frac{\Phi
_{,r}}{%
2\Phi }(\gamma ^2+1)+\frac{\Delta _{,r}}{2\Delta }(\gamma
^2-2)-\frac{f_{,r}}%
f(2\gamma ^2-1)-\frac 2r\right) \right]  \nonumber \\
&& \nonumber \\
&&-\frac{\Phi _{,r}}{2\Phi }\frac{\Sigma _{,r}}{2\Sigma }(1+\gamma
)^2(1-\gamma )^2+\frac{\Phi _{,rr}}{2\Phi }(1-\gamma ^2)+\frac{f_{,rr}}%
f\gamma ^2 \nonumber \\
&&  \nonumber \\
&&+\frac{f_{,r}}f\gamma ^2\left[ \frac{\Delta _{,r}}{2\Delta }(2-\gamma
^2)+(\gamma ^2-1)\left( \frac{\Sigma _{,r}}{2\Sigma
}+\frac{f_{,r}}f\right)
\right]  \nonumber \\
&&  \nonumber \\
&&+\frac 1r\left[ \frac{\Delta _{,r}}{2\Delta }(2-\gamma
^2)+\frac{\Sigma
_{,r}}{2\Sigma }(\gamma ^2-1)+\frac{f_{,r}}f(2\gamma ^2-1)\right]
\nonumber
\\
&&  \nonumber \\
&&\}  \nonumber \\
&&  \nonumber \\
&&+r^2\sin ^2\theta f^{2-4\gamma }\times  \nonumber \\
&&\{  \nonumber \\
&&\frac{\Phi _{,t}}{2\Phi }\left[ (1-\gamma ^2)\left( \frac{\Phi
_{,t}}{%
2\Phi }(1+\gamma ^2)+\frac{\Delta _{,t}}{2\Delta }(2-\gamma
^2)-\frac{f_{,t}}%
f(2\gamma ^2-4\gamma +1)\right) \right]  \nonumber \\
&& \nonumber \\
&&-\frac{\Phi _{,t}}{2\Phi }\frac{\Sigma _{,t}}{2\Sigma }(1+\gamma
)^2(1-\gamma )^2 \nonumber \\
&&  \nonumber \\
&&+\frac{\Delta _{,t}}{2\Delta }(\gamma ^2-2)\left[ (4-\gamma ^2)\frac{%
\Delta _{,t}}{2\Delta }+\frac{\Sigma _{,t}}\Sigma (\gamma ^2-1)\right]
+\left( \frac{\Sigma _{,t}}{2\Sigma }\right) ^2(1-\gamma ^4)  \nonumber
\\
&&  \nonumber \\
&&+\frac{f_{,t}}f\left[ \frac{f_{,t}}f\gamma (2-\gamma )(\gamma
-1)^2+(\gamma ^2-4\gamma +2)\left( \frac{\Delta _{,t}}{2\Delta
}(2-\gamma
^2)-\frac{\Sigma _{,t}}{2\Sigma }(1-\gamma ^2)\right) \right]  \nonumber
\\
&&  \nonumber \\
&&-\frac{f_{,tt}}f(1-\gamma )^2+\frac{\Delta _{,tt}}{2\Delta }(2-\gamma
^2)+(\gamma ^2-1)\left[ \frac{\Sigma _{,tt}}{2\Sigma }+\frac{\Phi
_{,tt}}{%
2\Phi }\right]  \nonumber \\
&&\}  \nonumber \\
&&  \nonumber \\
&&+\sin ^2\theta f^{2(1-\gamma ^2)}\Phi ^{\gamma ^2-1}\times \nonumber
\\
&&  \nonumber \\
&&\left[ \frac{\Sigma _{,\theta \theta }}{2\Sigma }(1-\gamma ^2)+\frac{%
\Sigma _{,\theta }}{2\Sigma }\left( \frac{\Sigma _{,\theta }}{2\Sigma }%
(\gamma ^4-1)-\frac{\Phi _{,\theta }}{2\Phi }(1+\gamma )^2(1-\gamma
)^2\right) \right]   \label{29}
\end{eqnarray}

\begin{eqnarray}
G_{tt} &=&\frac{\Sigma _{,t}}{2\Sigma }(1-\gamma )^2\left[ \frac{\Phi
_{,t}}{%
2\Phi }(1+\gamma )^2+\frac{f_{,t}}f(1-\gamma ^2)\right]  \nonumber \\
&& \nonumber \\
&&+\frac{\Delta _{,t}}{2\Delta }(\gamma ^2-2)\left[ \frac{\Phi
_{,t}}{2\Phi }%
(1-\gamma ^2)+\frac{f_{,t}}f(1-\gamma )^2\right] \nonumber \\
&&  \nonumber \\
&&+\frac{f_{,t}}f(1-\gamma )^2\left[ \frac{\Phi _{,t}}{2\Phi }(2+2\gamma
)+%
\frac{f_{,t}}f(1-2\gamma )\right] +f^{4\gamma -2}\Delta ^{2-\gamma
^2}\Sigma
^{\gamma ^2-1}\times  \nonumber \\
&&\{  \nonumber \\
&&\frac{\Delta _{,r}}{2\Delta }(\gamma ^2-2)\left[
\frac{f_{,r}}f(1-\gamma
)^2+\frac{\Phi _{,r}}{2\Phi }\left( 1-\gamma ^2\right) \right]
\nonumber \\
&& \nonumber \\
&&+\frac{\Sigma _{,r}}{2\Sigma }(1-\gamma )^2\left[
\frac{f_{,r}}f(1-\gamma
^2)+\frac{\Phi _{,r}}{2\Phi }(1+\gamma )^2\right] \nonumber \\
&&  \nonumber \\
&&+\frac{f_{,r}}f(1-\gamma )^2\left[ \frac{f_{,r}}f(1-\gamma
^2)+\frac{\Phi
_{,r}}\Phi \gamma (1+\gamma )\right] +(1-\gamma ^4)\left( \frac{\Phi
_{,r}}{%
2\Phi }\right) ^2  \nonumber \\
&&  \nonumber \\
&&+\frac{\Phi _{,rr}}{2\Phi }(\gamma ^2-1)-\frac{f_{,rr}}f(1-\gamma
)^2+\frac 1r\left[ \frac{\Delta _{,r}}\Delta (\gamma ^2-2)+\frac{\Sigma
_{,r}%
}\Sigma (1-\gamma ^2)\right]  \nonumber \\
&&  \nonumber \\
&&+\frac 1r\left[ \frac{f_{,r}}f(3\gamma -1)(1-\gamma )+3\frac{\Phi
_{,r}}{%
2\Phi }(\gamma ^2-1)-\frac 1r\right]  \nonumber \\
&&\}  \nonumber \\
&&  \nonumber \\
&&+r^{-2}f^{2\gamma (2-\gamma )}\Phi ^{\gamma ^2-1}\times  \nonumber \\
&&\{  \nonumber \\
&&\frac{\Sigma _{,\theta \theta }}{2\Sigma }(\gamma ^2-1)+\frac{\Sigma
_{,\theta }}{2\Sigma }\left[ \frac{\Phi _{,\theta }}{2\Phi }(1-\gamma
)^2(1+\gamma )^2+\frac{\Sigma _{,\theta }}{2\Sigma }(1-\gamma ^4)\right]
\nonumber \\
&&  \nonumber \\
&&+(1-\gamma ^2)\left[ \frac{\Phi _{,\theta }}{2\Phi }-\frac{\Sigma
_{,\theta }}{2\Sigma }\right] \cot \theta +1  \nonumber \\
&&\}  \label{30}
\end{eqnarray}
where a comma denote partial derivation.

\section{Energy density for $\gamma =1+\varepsilon $}

Since we are interested in the effect of small deviations from spherical
symmetry, we shall assume that
\begin{equation}
\gamma =1+\varepsilon ,  \label{54}
\end{equation}
where $\varepsilon $ is a small constant that may be positive or
negative.
We restrict ourselves to first order in $\varepsilon ,$ so we shall
neglect
terms of order ${\cal O}(\varepsilon ^2)$ and higher.

Then, the $tt$ component of Einstein equations reads

\begin{eqnarray}
8\pi \rho &=&f^{2\varepsilon}\Delta^{1-2\varepsilon}\Sigma^{2\varepsilon}
\times \nonumber \\
&&\{ \nonumber \\
&& \varepsilon \frac{\Delta_{,r}}{2\Delta}
\frac{\Phi_{,r}}{\Phi} - \varepsilon \left( \frac{\Phi_{,r}}{\Phi}\right)^2
+ \varepsilon \frac{\Phi_{,rr}}{\Phi} \nonumber \\
&&+\frac{1}{r}\left( (2\varepsilon-1)\frac{\Delta_{,r}}{\Delta} -
2\varepsilon \left[ \frac{\Sigma_{,r}}{\Sigma} + \frac{f_{,r}}{f}\right] +
3\varepsilon \frac{\Phi_{,r}}{\Phi} - \frac{1}{r} \right) \nonumber \\
&&\}  \nonumber \\
&&+ f^{-2\varepsilon}\Phi^{2\varepsilon} \times \left [
\varepsilon\left(\frac{\Sigma_{,\theta\theta}}{\Sigma}-
\left[\frac{\Sigma_{,\theta}}{\Sigma}\right]^2- \cot{\theta}\left[
\frac{\Phi_{,\theta}}{\Phi}-\frac{\Sigma_{,\theta}}{\Sigma}\right]\right)
+1 \right] \label{54b}
\end{eqnarray}

\subsection{Schwarzschild-type solution}

Using (\ref{12}-\ref{17}) into (\ref{54b}) it is easy to show that, for
 $r\rightarrow 0$,
the energy density can be expressed as

\begin{equation}
\rho = \frac{m}{2\pi a^3}f^{2\varepsilon}\left[ \frac{3}{2}+
\varepsilon \left( 1+ \frac{1}{3\sqrt{1-\frac{2m}{a}}-1}\right)
\right]
\label{schdensid}
\end{equation}

\subsection{Adler-type solution}

In this case the energy density for $r\rightarrow 0$ is given, by means of
(\ref{18}-\ref{21}) and (\ref{54b}), by expression

\begin{equation}
\rho = \frac{3m}{4\pi a^3 \left( 1- 2m/a\right)^{1/3}}
\left(\frac{1-5m/2a}{\sqrt{1-2m/a}}\right)^{2\varepsilon-2/3}
\left[ \left(1-m/a\right)^{2/3}+
\frac{\varepsilon}{\left(1-5m/2a\right)^{1/3}}\right]
\label{adldensid}
\end{equation}

\section{Conservation equations}

The left hand side of conservation equations
\[
T_{\nu ;\mu }^\mu =0
\]
are
\begin{eqnarray}
T_{t;\mu }^\mu &=&\Gamma ^2\times  \nonumber \\
&&\{  \nonumber \\
&&\left[ \frac{f_{,t}}f\left( \gamma -1\right) \left( \gamma -2\right)
+%
\frac{\Delta _{,t}}{2\Delta }\left( \gamma ^2-2\right) +\left(
\frac{\Phi
_{,t}}{2\Phi }+\frac{\Sigma _{,t}}{2\Sigma }\right) \left( 1-\gamma
^2\right) +\frac{2\Gamma _{,t}}\Gamma \right] \times  \nonumber \\
&&\left( \rho +p_{xx}w_x^2+p_{yy}w_y^2+2p_{xy}w_xw_y\right)  \nonumber
\\
&&  \nonumber \\
&&+\rho _{,t}+p_{xx,t}w_x^2+2p_{xx}w_xw_{x,t}+p_{yy,t}w_y^2  \nonumber
\\
&&  \nonumber \\
&&+2p_{yy}w_yw_{y,t}+2p_{xy,t}w_xw_y+2p_{xy}\left(
w_xw_{y,t}+w_{x,t}w_y\right)  \nonumber \\
&&\}  \nonumber \\
&&  \nonumber \\
&&+f^{2\gamma -1}\Delta ^{1-\gamma ^2/2}\Sigma ^{(\gamma ^2-1)/2}\Gamma
\times  \nonumber \\
&&\{  \nonumber \\
&&\left[ \frac{f_{,r}}f\left( \gamma ^2+1\right) +\frac{\Phi
_{,r}}{2\Phi }%
\left( 1-\gamma ^2\right) +\frac{\Gamma _{,r}}\Gamma +\frac 2r\right]
\times
\nonumber \\
&&\left( \Gamma w_x\rho +p_{xx}w_x\Lambda _x+p_{yy}w_y\Lambda
+p_{xy}\left[
w_x\Lambda +w_y\Lambda _x\right] \right)  \nonumber \\
&&  \nonumber \\
&&+\Gamma _{,r}w_x\rho +\Gamma w_{x,r}\rho +\Gamma w_x\rho
_{,r}+p_{xx,r}w_x\Lambda _x  \nonumber \\
&&  \nonumber \\
&&+p_{yy,r}w_y\Lambda +p_{xy,r}\left[ w_x\Lambda +w_y\Lambda _x\right]
\nonumber \\
&&  \nonumber \\
&&+w_{x,r}\left[ p_{xx}\Lambda _x+p_{xy}\Lambda \right] +w_{y,r}\left[
p_{yy}\Lambda +p_{xy}\Lambda _x\right]  \nonumber \\
&&  \nonumber \\
&&+\Lambda _{x,r}\left[ p_{xx}w_x+p_{xy}w_y\right] +\Lambda _{,r}\left[
p_{yy}w_y+p_{xy}w_x\right]  \nonumber \\
&&\}  \nonumber \\
&&  \nonumber \\
&&+r^{-1}f^{2\gamma -\gamma ^2}\Phi ^{(\gamma ^2-1)/2}\Gamma \times
\nonumber \\
&&\{  \nonumber \\
&&\left[ \frac{\Sigma _{,\theta }}{2\Sigma }\left( 1-\gamma ^2\right)
+\frac{%
\Gamma _{,\theta }}\Gamma +\cot \theta \right] \times  \nonumber \\
&&\left( \Gamma w_y\rho +p_{xx}w_x\Lambda +p_{yy}w_y\Lambda
_y+p_{xy}\left[
w_y\Lambda +w_x\Lambda _y\right] \right)  \nonumber \\
&&  \nonumber \\
&&+\Gamma _{,\theta }w_y\rho +w_{y,\theta }\left( \rho \Gamma
+p_{yy}\Lambda
_y+p_{xy}\Lambda \right) +w_{x,\theta }\left( p_{xx}\Lambda
+p_{xy}\Lambda
_y\right)  \nonumber \\
&&  \nonumber \\
&&+\Lambda _{,\theta }\left( p_{xx}w_x+p_{xy}w_y\right) +\Lambda
_{y,\theta
}\left( p_{yy}w_y+p_{xy}w_x\right)  \nonumber \\
&&  \nonumber \\
&&+\rho _{,\theta }\Gamma w_y+p_{xx,\theta }w_x\Lambda +p_{yy,\theta
}w_y\Lambda _y+p_{xy,\theta }\left[ w_y\Lambda +w_x\Lambda _y\right]
\nonumber \\
&&\}  \nonumber \\
&&  \nonumber \\
&&+\left[ \frac{f_{,t}}f\gamma \left( \gamma -1\right) +\frac{\Phi
_{,t}}{%
2\Phi }\left( 1-\gamma ^2\right) \right] \times  \nonumber \\
&&\left( \Gamma ^2w_y^2\rho +p_{xx}\Lambda ^2+p_{yy}\Lambda
_y^2+2p_{xy}\Lambda \Lambda _y\right)  \nonumber \\
&&  \nonumber \\
&&+\frac{f_{,t}}f\left( 1-\gamma \right) p_{zz}  \nonumber \\
&&  \nonumber \\
&&+\left[ \frac{f_{,t}}f\left( 1-\gamma \right) +\frac{\Delta
_{,t}}{2\Delta
}\left( \gamma ^2-2\right) +\frac{\Sigma _{,t}}{2\Sigma }\left( 1-\gamma
^2\right) \right] \times  \nonumber \\
&&\left( \Gamma ^2w_x^2\rho +p_{xx}\Lambda _x^2+p_{yy}\Lambda
^2+2p_{xy}\Lambda \Lambda _x\right)  \label{A1}
\end{eqnarray}

\begin{eqnarray}
T_{r;\mu }^\mu &=&-f^{1-2\gamma }\Delta ^{-1+\gamma ^2/2}\Sigma
^{(1-\gamma
^2)/2}\Gamma \times  \nonumber \\
&&\{  \nonumber \\
&&\left[ \frac{f_{,t}}f\left( \gamma -1\right) \left( \gamma -3\right)
+%
\frac{\Delta _{,t}}\Delta \left( \gamma ^2-2\right) +\left( \frac{\Phi
_{,t}%
}{2\Phi }+\frac{\Sigma _{,t}}\Sigma \right) \left( 1-\gamma ^2\right)
+\frac{%
\Gamma _{,t}}\Gamma \right] \times  \nonumber \\
&&\left( \Gamma w_x\rho +p_{xx}w_x\Lambda _x+p_{yy}w_y\Lambda
+p_{xy}\left[
w_x\Lambda +w_y\Lambda _x\right] \right)  \nonumber \\
&&  \nonumber \\
&&+\Gamma w_x\rho _{,t}+\Gamma _{,t}w_x\rho +p_{xx,t}w_x\Lambda
_x+p_{yy,t}w_y\Lambda +p_{xy,t}\left[ w_x\Lambda +w_y\Lambda _x\right]
\nonumber \\
&&  \nonumber \\
&&+w_{x,t}\left( \Gamma \rho +p_{xx}\Lambda _x+p_{xy}\Lambda \right)
+w_{y,t}\left( p_{yy}\Lambda +p_{xy}\Lambda _x\right)  \nonumber \\
&&  \nonumber \\
&&+\Lambda _{x,t}\left( p_{xy}w_y+p_{xx}w_x\right) +\Lambda _{,t}\left(
p_{yy}w_y+p_{xy}w_x\right)  \nonumber \\
&&\}  \nonumber \\
&&  \nonumber \\
&&-\left[ \frac{f_{,r}}f\left( \gamma ^2-\gamma +1\right) +\frac{\Phi
_{,r}}{%
2\Phi }\left( 1-\gamma ^2\right) +\frac 2r\right] \times  \nonumber \\
&&\left( \Gamma ^2w_x^2\rho +p_{xx}\Lambda _x^2+p_{yy}\Lambda
^2+2p_{xy}\Lambda \Lambda _x\right)  \nonumber \\
&&  \nonumber \\
&&-\left[ 2\Gamma \Gamma _{,r}w_x^2\rho +2\Gamma ^2w_xw_{x,r}\rho
+\Gamma
^2w_x^2\rho _{,r}+p_{xx,r}\Lambda _x^2+p_{yy,r}\Lambda
^2+2p_{xy,r}\Lambda
\Lambda _x\right]  \nonumber \\
&&  \nonumber \\
&&-2\Lambda _{x,r}\left[ p_{xx}\Lambda _x+p_{xy}\Lambda \right]
-2\Lambda
_{,r}\left[ p_{xy}\Lambda _x+p_{yy}\Lambda \right]  \nonumber \\
&&  \nonumber \\
&&-r^{-1}f^{1-\gamma ^2}\Delta ^{-1+\gamma ^2/2}\Phi ^{(\gamma
^2-1)/2}\Sigma ^{(1-\gamma ^2)/2}\times  \nonumber \\
&&\{  \nonumber \\
&&\left[ \frac{\Sigma _{,\theta }}\Sigma \left( 1-\gamma ^2\right) +\cot
\theta \right] \times  \nonumber \\
&&\left( \Gamma ^2w_xw_y\rho +p_{xx}\Lambda _x\Lambda +p_{yy}\Lambda
_y\Lambda +p_{xy}\left[ \Lambda ^2+\Lambda _x\Lambda _y\right] \right)
\nonumber \\
&&  \nonumber \\
&&+\Gamma ^2w_xw_y\rho _{,\theta }+p_{xx,\theta }\Lambda _x\Lambda
+p_{yy,\theta }\Lambda _y\Lambda +p_{xy,\theta }\left[ \Lambda
^2+\Lambda
_x\Lambda _y\right]  \nonumber \\
&&  \nonumber \\
&&+2\Gamma \Gamma _{,\theta}w_xw_y\rho +\Gamma ^2w_{x,\theta }w_y\rho +\Gamma
^2w_xw_{y,\theta }\rho +\Lambda _{x,\theta }\left[ p_{xx}\Lambda
+p_{xy}\Lambda _y\right]  \nonumber \\
&&  \nonumber \\
&&+\Lambda _{y,\theta }\left[ p_{yy}\Lambda +p_{xy}\Lambda _x\right]
+\Lambda _{,\theta }\left[ 2p_{xy}\Lambda +p_{xx}\Lambda
_x+p_{yy}\Lambda
_y\right]  \nonumber \\
&&\}  \nonumber \\
&&  \nonumber \\
&&-\frac{f_{,r}}f\gamma \Gamma ^2\left( \rho
+p_{xx}w_x^2+p_{yy}w_y^2+2p_{xy}w_xw_y\right)  \nonumber \\
&&  \nonumber \\
&&+\left[ \frac 1r+\frac{f_{,r}}f\left( 1-\gamma \right) \right] p_{zz}
\nonumber \\
&&  \nonumber \\
&&+\left[ \frac 1r+\frac{f_{,r}}f\gamma \left( \gamma -1\right)
+\frac{\Phi
_{,r}}{2\Phi }\left( 1-\gamma ^2\right) \right] \times  \nonumber \\
&&\left( \Gamma ^2w_y^2\rho +p_{xx}\Lambda ^2+p_{yy}\Lambda
_y^2+2p_{xy}\Lambda \Lambda _y\right)  \label{A2}
\end{eqnarray}

\begin{eqnarray}
T_{\theta ;\mu }^\mu &=&-rf^{\gamma ^2-2\gamma }\Phi ^{(1-\gamma
^2)/2}\Gamma \times  \nonumber \\
&&\{  \nonumber \\
&&\left[ \frac{2f_{,t}}f\left( \gamma -1\right) ^2+\frac{\Delta _{,t}}{%
2\Delta }\left( \gamma ^2-2\right) +\left( \frac{\Sigma _{,t}}{2\Sigma
}+%
\frac{\Phi _{,t}}\Phi \right) \left( 1-\gamma ^2\right) +\frac{\Gamma
_{,t}}%
\Gamma \right] \times  \nonumber \\
&&\left( \Gamma w_y\rho +p_{xx}w_x\Lambda +p_{yy}w_y\Lambda
_y+p_{xy}\left[
w_y\Lambda +w_x\Lambda _y\right] \right)  \nonumber \\
&&  \nonumber \\
&&+\Gamma w_y\rho _{,t}+p_{xx,t}w_x\Lambda +p_{yy,t}w_y\Lambda
_y+p_{xy,t}\left[ w_y\Lambda +w_x\Lambda _y\right]  \nonumber \\
&&  \nonumber \\
&&+\Gamma _{,t}w_y\rho +\Lambda_{,t}\left[ p_{xx}w_x+p_{xy}w_y\right]
+ \Lambda_{y,t}\left[ p_{yy}w_y+p_{xy}w_x\right]  \nonumber \\
&&  \nonumber \\
&&+w_{y,t}\left[ p_{yy}\Lambda _y+p_{xy}\Lambda +\Gamma \rho \right]
+w_{x,t}\left[ p_{xx}\Lambda +p_{xy}\Lambda _y\right]  \nonumber \\
&&\}  \nonumber \\
&&  \nonumber \\
&&-rf^{\gamma ^2-1}\Delta ^{1-\gamma ^2/2}\Phi ^{(1-\gamma ^2)/2}\Sigma
^{(\gamma ^2-1)/2}\times  \nonumber \\
&&\{  \nonumber \\
&&\left[ \frac{f_{,r}}f\left( 2\gamma ^2-2\gamma +1\right) +\frac{\Phi
_{,r}}%
\Phi \left( 1-\gamma ^2\right) +\frac 3r\right] \times  \nonumber \\
&&\left( \Gamma ^2w_xw_y\rho +p_{xx}\Lambda _x\Lambda +p_{yy}\Lambda
_y\Lambda +p_{xy}\left[ \Lambda ^2+\Lambda _x\Lambda _y\right] \right)
\nonumber \\
&&  \nonumber \\
&&+\Gamma ^2w_xw_y\rho _{,r}+p_{xx,r}\Lambda _x\Lambda +p_{yy,r}\Lambda
_y\Lambda +p_{xy,r}\left[ \Lambda ^2+\Lambda _x\Lambda _y\right]
\nonumber
\\
&&  \nonumber \\
&&+2\Gamma \Gamma _{,r}w_xw_y\rho +\Gamma ^2w_{x,r}w_y\rho +\Gamma
^2w_xw_{y,r}\rho  \nonumber \\
&&  \nonumber \\
&&+\Lambda _{,r}\left[ p_{xx}\Lambda _x+p_{yy}\Lambda _y+2p_{xy}\Lambda
\right]  \nonumber \\
&&  \nonumber \\
&&+\Lambda _{y,r}\left[ p_{yy}\Lambda +p_{xy}\Lambda _x\right] +\Lambda
_{x,r}\left[ p_{xx}\Lambda +p_{xy}\Lambda _y\right]  \nonumber \\
&&\}  \nonumber \\
&&  \nonumber \\
&&-\left[ \frac{\Sigma _{,\theta }}{2\Sigma }\left( 1-\gamma ^2\right)
+\cot
\theta \right] \left( \Gamma ^2w_y^2\rho +p_{xx}\Lambda ^2+p_{yy}\Lambda
_y^2+2p_{xy}\Lambda \Lambda _y\right)  \nonumber \\
&&  \nonumber \\
&&+p_{zz}\cot \theta  \nonumber \\
&&  \nonumber \\
&&-\left[ \Gamma ^2w_y^2\rho _{,\theta }+p_{xx,\theta }\Lambda
^2+p_{yy,\theta }\Lambda _y^2+2p_{xy,\theta }\Lambda \Lambda _y\right]
\nonumber \\
&&  \nonumber \\
&&-\left[ 2\Gamma \Gamma _{,\theta }w_y^2\rho +2\Gamma ^2w_yw_{y,\theta
}\rho \right]  \nonumber \\
&&  \nonumber \\
&&-2\Lambda _{y,\theta }\left[ p_{yy}\Lambda _y+p_{xy}\Lambda \right]
-2\Lambda _{,\theta }\left[ p_{xx}\Lambda +p_{xy}\Lambda _y\right]
\nonumber
\\
&&  \nonumber \\
&&+\frac{\Sigma _{,\theta }}{2\Sigma }\left( 1-\gamma ^2\right) \left(
\Gamma ^2w_x^2\rho +p_{xx}\Lambda _x^2+p_{yy}\Lambda ^2+2p_{xy}\Lambda
\Lambda _x\right)  \label{A3}
\end{eqnarray}

\begin{equation}
T_{\varphi ;\mu }^\mu =-p_{zz,\varphi }  \label{A4}
\end{equation}

\end{document}